%Paper: astro-ph/9510014
%From: SYLOS@roma1.infn.it
%Date: Wed, 4 Oct 1995 13:04:28 +0100 (WET)

%===================================================================
%Astrophysical Letters and Communications
%Editor in Chief
%Giorgio G.  C.  Palumbo
%Dipartimento di Astronomia
%Universita` degli Studi di Bologna
%via Zamboni 33
%40126 Bologna
%Italy
%e-mail:  ggcpalumbo@astbo3.bo.astro.it
%phone:  Italy+51+259424

%------------------macro-tex-instructions-----------------------

%------------------begin definitions----------------------------

\def   \ni {\noindent}

\def   \cl {\centerline}

\def   \hm {h^{-1}Mpc}

\def   \etal {{\it et al.}}

\def   \lam {\lambda}

\def   \rapj {ApJ }

\def   \rmnras {MNRAS }

\def   \ssk {\vskip  5truept}

\def   \bsk {\vskip 15truept}

\def   \newline {\hfil\break}

%------------end definitions-----------------------------------------

%\input psfig.tex

\magnification=1000

\hsize 5truein

\vsize 8truein

\font\abstract=cmr8

\font\text=cmr10

\font\affiliation=cmssi10

\font\author=cmss10

\font\title=cmssbx10 scaled\magstep2

\def\ref{\par\noindent\hangindent 15pt}

\nopagenumbers

\null

\vskip 3.0truecm

\baselineskip = 12pt

{\title  Power spectrum
for fractal distributions
\ni

}

                                 %% end of font "title"

\bsk \bsk

{\author  F. SYLOS LABINI$^{1,2}$
and L. AMENDOLA$^{3}$

     %% beginning of font "author"

\ni

}

                                 %% end of font "author"

\bsk

{\affiliation                   %% beginning of font "affiliation"

(1) Dipartimento di Fisica, Universita di Bologna, Italy

(2) Dipartimento di Fisica, Universita di Roma "La Sapienza", 00185 Roma, Italy

(3) Osservatorio Astronomico di Roma, Viale del Parco Mellini 84,
I-00136 Roma, Italy

}
                       %% end of font "affiliation"

\bsk

\cl {\it (Received July 1995)}

\bsk

\baselineskip = 9pt

{\abstract
We study the behaviour of the power
spectrum (PS) in the case of fractal structures. We show that
in this case the main observational features of the PS,
the  large scale flattening and the scaling of the
amplitude with  sample depth,
are related to finite size effects, due to the fractal
nature of galaxy distribution in the sample. Comparing with the
recent results of the PS for the CfA2 we conclude that
this catalog is consistent with the fractal description.
\ni

}

\bsk

\baselineskip = 12pt

{\text
                            %% beginning of font "text"

%\ni 1. INTRODUCTION
%\ssk

Identifying the scale at which our Universe becomes homogeneous,
if any, is a crucial task of contemporary cosmology, and
a very debated one, especially for what concerns
the homogeneity of the luminous matter at the
present.
 Essentially all the currently
elaborated models of galaxy formation
 assume large scale homogeneity and
predict that the galaxy
power spectrum (PS)
decreases both toward small scales and toward large
scales, with a turnaround  at some scale $\lambda_f$
that can be taken as separating ``small'' from ``large'' scales.
Then, because of the assumption of homogeneity, the power spectrum amplitude
should be independent of the survey scale, any residual
variation being attributed to luminosity bias (or to the
fact that the survey scale has not yet reached the homogeneity scale).
However,
the crucial clue to this
picture, the firm determination of the
scale $\lambda_f$, is still missing, although
some surveys do indeed produce a  turnaround
 scale around 100 $\hm$
(Baugh \& Efstathiou 1992;
Feldman \etal 1994). Recently, the CfA2  survey
analyzed by  Park \etal (1994; PVGH) (and confirmed by SSRS2
- Da Costa \etal (1994, DVGHP)), showed a $n=-2$ slope up to $\sim 30 \hm$,
a milder $n\approx -1$ slope  up to 200 $\hm$, and some tentative
indication of flattening on even larger scales. They also find
that deeper subsamples have higher power amplitude,
i.e. that the amplitude scales with the sample depth.

In this paper we argue  that both features, bending and scaling,
are a manifestation of
 the finiteness of the survey volume, and that they
cannot be
interpreted as the convergence to homogeneity, nor to a power spectrum
flattening.
The systematic effect of the survey finite size is in fact
 to suppress
power at large scale, mimicking a real flattening.
We show that
even a fractal distribution of matter, i.e. a distribution which
never reaches  homogeneity, shows a sharp flattening in the PS
(even  when the correction proposed by Peacock \& Nicholson (1991) is
applied to the data), and that its  amplitude
depends on the survey size  (see Sylos Labini \& Amendola, 1995).
The reason of such behavior is that
the standard  power spectrum (hereafter SPS)
measures directly the contributions of different scales to the galaxy
density contrast $ \delta\rho/\rho$.
It is clear then
that the density contrast, and all the quantities based on it,
is meaningful only when one can define
a constant density, i.e. reliably identify
the sample density with
the average density of all the Universe.
When this is not true, and we argue that is
indeed an incorrect assumption
in all the cases investigated so far, a false interpretation of the results may
occur, since both
the shape and the amplitude of the power spectrum depend on the
survey size.

%\bsk
%\ni 2. THE STANDARD POWER SPECTRUM
%\ssk

Let us recall the basic  notation of the power spectrum analysis.
Following Peebles (1980)  we imagine that the Universe is periodic
in a volume $ V_{u}$, with $ V_{u}$ much
larger than the (assumed) maximum
correlation length. The survey volume $V\in V_u$
 contains $ N$ galaxies at positions $ \vec{r_i}$,
  and the galaxy density contrast is
$
%%\label{e4}
\delta(\vec{r}) = [n(\vec{r})/\hat n] -1
$
where it is assumed that exists a
well defined constant density $\hat n$, obtained
averaging over a sufficiently large scale.
The density function
%$ n(\vec{r})$
can be described by a sum of delta functions:
$~n(\vec{r}) = \sum_{i=1}^{N} \delta^{(3)} (\vec{r}-\vec{r_{i}})\,.~$
Expanding the density contrast   in its Fourier components we have
$$
%\label{e7}
\delta_{\vec{k}} = 1/N \sum _{j \epsilon V}
e^{i\vec{k}\vec{r_{j}}} - W(\vec{k})\,,\eqno{1}
$$
where
%$$
%%\label{e7b}
$~W(\vec{k}) = V^{-1} \int d{\vec{r}} W(\vec{r})
 e^{i\vec{k}\vec{r}}\,~$
%$$
is the Fourier transform of the survey window $W(\vec{r})$,
defined to be unity inside the survey region, and zero outside.
If $\xi(\vec{r})$ is the correlation function of the galaxies,
($\xi(\vec{r}) = <n(\vec{r})n(0)>/\hat n^2 -1$)
the true PS $ P(\vec{k})$ is defined as
the Fourier conjugate of the
correlation function $\xi(r)$.
Because of isotropy the PS can be simplified to
$$
%\label{e12}
P(k)  =4\pi \int \xi(r)  \sin(kr)/(kr) r^{2}dr\,.\eqno{2}
$$
The  variance of $ \delta_{\vec{k}}$
is (Peebles 1980)
%Peacock \& Nicholson 1991; Fisher et al. 1993):
%$$
%%\label{e9}
$<|\delta_{\vec{k}}|^{2}> = N^{-1}+ V^{-1}\tilde P(\vec{k})\,.
$
%$$
i.e. the sum of a shot noise term and of
the true PS convolved with a window function
 (Sylos Labini \& Amendola, 1995)

We apply now this standard analysis to a fractal distribution.
Consider a self-similar system, where the number of points inside
a certain radius $ r$  scales
according to the mass-length relation
(Mandelbrot, 1982)
%$$
%%\label{e14}
$~N(r) = Br^{D}\,,~$
%$$
with $ D<3$ (the case $ D=3$ corresponds to the
homogenous distribution). Then we have the following correlation function
$$
%\label{e16}
\xi(r) = [(3-\gamma)/3](r/R_{s})^{-\gamma} -1\,,\eqno{3}
$$
 where  $ \gamma=3-D$,
On scales larger that $R_s$ the $\xi(r)$ cannot be calculated without
making assumptions on the distribution outside the sampling volume
(in particular, assumption of homogeneity),
which is just what we want to avoid.
When the survey volume is not spherical, the scale $R_s$ is
 of the order of the largest sphere completely contained inside
the survey.
Both
the amplitude and the shape of $\xi(r)$ are
therefore scale-dependent  in the
case of a fractal distribution (CP92).
inside a sphere of radius $R_s$ turns out to be
$$
%\label{e19}
P(k) =
%\int^{R_{s}}_{0}
%   4\pi \sin(kr)/(kr) \left[ (3-\gamma)/(3)
%\left((r)/(R_{s})\right)^{-\gamma} -1\right] r^{2}dr=
(a(k,R_s) R_{s}^{3-D})/(k^{D})-(b(k,R_s))/(k^{3})\,.\eqno{4}
$$
Notice that the integral has to be evaluated inside $R_s$
because we want to compare $P(k)$  with its {\it estimation}
 in a finite size spherical survey of scale $R_s$.
 In the  general case, we must deconvolve the
 window contribution
 from $P(k)$; $R_s$ is then a characteristic window scale.
%Eq. (\ref{e19})
The previous equation
 shows the two scale-dependent features of the PS. First,
the amplitude of the PS
depends on the sample depth.
Secondly,
the shape of the PS
is characterized by two  scaling regimes:
the first one, at high wavenumbers,
is related to the real fractal dimension,
while the second one arises only because of
the finiteness of the sample.
In the case of $ D=2$
%in eq.\ref{e19}
one has:
%$$
%%\label{e22}
$~a = (4\pi)/(3) (2+\cos(kR_{s}))\,,~$ and
%$$
% NO $R_MIN$  MA 0
%$$
%%\label{e23}
$~b = 4\pi \sin (kR_{s})\,.~$
%$$
The PS is then a power-law with exponent
$ -2$ at high wavenumbers,
it flattens at low wavenumbers and reaches a maximum at
$k\approx 4.3/R_s$, i.e. at a scale $\lam\approx 1.45 R_s$.
%The scale at which the transition occurs
%is thus related to the sample depth.
In a real survey, things are complicated by the window function,
so that the flattening (and the turnaround) scale can only be determined
numerically.
%We study this behaviour in detail in Sec.  3.

To avoid the mean density normalization, which gives
misleading results in fractals distributions,
%To this aim,
we
introduce now the scale-independent PS (SIPS) of the density $ \rho(\vec{r})$,
%a quantity that  does not involve the computation
%of the average density, and thus
a quantity which  gives
 an unambiguous information
of the statistical properties of the system.
We first introduce the density correlation function
%$$
%%\label{e24}
$G(\vec{r}) = <\rho(\vec{x}+\vec{r})\rho(\vec{x})> = A r^{-(3-D)}\,,
$
%$$
where the last equality holds in the case
of a fractal distribution with dimension {\it D},
and where {\it A} is a constant
%determined by the lower cut-offs of the distribution (CP92).
 Defining
the SIPS  as the Fourier conjugate of the correlation function $G(r)$,
%$$
%%\label{e30}
%\Pi(k)  = 4\pi \int G(r)  \frac{\sin(kr)}{kr} r^{2}dr\,,
%$$
one obtains that in a finite spherical volume
%$$
%%\label{e31}
$\Pi(k) \sim A' k^{-D}$,
%$$
(where $A'=4\pi(1-\cos(k R_s))$ if $D=2$)
so that the SIPS is a single power law extending all over the system size,
without  amplitude  scaling
 (except for $kR_s\ll 1$).
In analogy to the procedure above,
we consider the Fourier transform of the density
%$$
%%\label{e26}
$\rho_{\vec{k}} = V^{-1} \sum_{j\in V} e^{-i\vec{k}\vec{x_j}}\,,
$
%$$
and its variance
%$$
%%\label{e27}
$<|\rho_{\vec{k}}|^{2}> = V^{-1}\tilde \Pi(\vec{k})+ N^{-1} \,,
$
%$$
where $\tilde \Pi(\vec{k})$ is the same
%as in Eq. (\ref{e9b}),
of the previous case but
with $<|\rho_{\vec{k'}}|^{2}>$ instead of $<|\delta_{\vec{k'}}|^{2}>$.

%\bsk

%\ni 3. RESULTS AND CONCLUSION
%\ssk

%
In Fig.1 we show the observational results
of PVGH and compare them with the PS of a fractal
distribution with $D=2$,
 adopting the same
technique used for real redshift surveys (Peacock \& Nicholson 1991;
 Fisher et al. 1993; PVGH; DVGHP).
Let us  summarize the observational results of
PVGH, by confronting them with
 the PS for
a fractal distribution (Fig. 1):
i) for $ k \ge 0.25$ ($ \lambda \le 25 h^{-1}$ Mpc) the PS
in a volume limited sample
is very close to a power law with slope $ n=-2.1$. In our view, this is the
behaviour at high wavenumbers
connected with the real fractal dimension.
ii) For  $ 0.05\le k \le 0.2$  ($120\hm>\lambda>30 \hm$) and the spectrum
is less steep, with a slope  about $ -1.1$.
This bending is, in  our view, solely due to the
finite size of the sample.
iii) The  amplitude of the
volume limited subsample CfA2-130
PS is $ \sim 40\%$ larger
than for CfA2-101.
%; the same trend is found in the $ \xi(r)$ analysis.
This linear scaling of the amplitude can be
understood again considering that the sample is
fractal with $ D=2$.
The same behaviour has been found in the analysis of the
$ \xi(r)$:  $ r_{0}$
scales linearly with the sample depth according to Eq.3 (CP92).
%--------------------------  figure 1
\midinsert
\vskip 0.5truecm
\par\noindent
%\centerline{\psfig{figure=figledasesto1.post,height=12truecm,width=10true}}
\leftskip=1truecm
\rightskip=1truecm
\noindent
{\abstract
FIGURE 1.  Comparison of
 power spectra of fractal distribution (triangles)
 with the CfA2 survey (squares).
 In the top panel, we plot the PS of the subsample CfA2-130
 (PVGH) along with the PS of our artificial fractal distribution
 (without the error bars for clarity).
 In the bottom panel, we plot CfA2-101 (PVGH) and a subsample of
 the same fractal as above, with a correspondingly scaled depth.
 }
\vskip 0.5truecm
\leftskip=0truecm
\rightskip=0truecm
\endinsert
%---------------------------------
The authors (PVGH)
explain this fact considering the dependence
of galaxy clustering on luminosity: brighter galaxies
correlate more than fainter ones.
It is certainly possible that
both mechanisms, the luminosity segregation and
the intrinsic self-similarity of the distribution,
%  SHOULD BE
are correct, and
each one explains part of the scaling.
However, PVGH do {\it not}
 detect such a luminosity segregation for the two largest
subsamples, CfA101 and CfA130, to which we are comparing our analysis
 here.
It seems therefore that the amplitude scaling at these scales can be entirely
attributed to the fractal scaling.
Our conclusion is then that the fractal nature of
the galaxy distribution can explain, to the scales
surveyed so far, the shift of the amplitude
with sample depth of the PS and of $\xi(r)$.

Finally we stress that the fractal dimension of
the galaxy clustering rises from
 $ D\sim1.4$ for CfA1 to $ D\sim 2$ for CfA2,
in agreement with the result of other independent surveys:
Perseus-Pisces (Guzzo \etal 1992; Sylos Labini \etal 1995),
and  ESP
(Pietronero \& Sylos Labini 1994, Baryshev et al., 1994).
\bsk

\ni Acknowledgements:

We wish to thank L. Pietronero for very useful discussions.

\bsk

\vskip 0.1truecm
\ni {REFERENCES}

\ref  Baryshev, Yu. V., Sylos Labini, F.,
Montuori, M., Pietronero, L.  1994,
 {\it  Vistas in Astron.} {\bf 38}, Part 4, 419

\ref     Coleman, P. H. \& Pietronero, L. 1992,
  Phys. Rep.  213, 311.

\ref   Da Costa et al. 1994,
 \rapj  424, L1 (DPVGH).

\ref Davis, M., Peebles, P. J. E. 1983,
 \rapj 267,465

\ref   Fisher et al. 1993,
 \rapj  402, 42

\ref   Guzzo, L., Iovino, A.,
Chincarini, G., Giovanelli, R.,
Haynes,M.  1992,
 \rapj  382, L5.

%\ref   Itoh M., Suginohara T. \& Suto Y. 1992, PASJ 44, 481.

\ref   Mandelbrot, B. B.  1982, The fractal geometry of Nature,
 (New York: W. H. Freeman).

\ref  Park, C., Vogeley, M.S., Geller, M.,\&  Huchra, J.
 1994
 \rapj  431, 569 (PVGH)

\ref  Peacock, J.A., Nicholson, D. 1991,
 \rmnras 235, 307

\ref  Peebles, P. J. E. 1980,
The Large-Scale Structure
of the Universe (Princeton:
Princeton Univ. Press)

\ref   Pietronero, L., Sylos Labini, F. 1995,
  in {\it Birth of the Universe and fundamental physics},
F. Occhionero ed., (Berlin: Springer Verlag) in press

\ref   Sylos Labini, F., M. Montuori,  Pietronero, L. 1995
 preprint

\ref   Sylos Labini, F. \& L. Amendola, 1995 preprint

\ssk

}
%% end of the font "text"
\end